\pdfoutput=1 
\documentclass{JINST}
\usepackage{tabularx}
\usepackage{acronym}
\usepackage{subfigure}
\usepackage{graphicx}
\usepackage[english]{babel}
\usepackage{subfloat}
\usepackage{cite}
\usepackage{caption}
\usepackage{lineno}
 \usepackage{amsmath}
\usepackage{subfloat}

\makeatletter
\let\@fnsymbol\@arabic
\makeatother

\let\ifpdf\relax

\title{The design of the ILD forward region\footnote{Talk presented at the International Workshop on Future Linear Colliders, 
LCWS 2016, 5 - 9 December 2016,  Morioka, Iwate, Japan.}\\ 
\small{\centerline{}}}

\author{ 

Aharon Levy$^a$\\
\centerline {\bf{\Large{\rm{On behalf of the FCAL collaboration}}}}\\ \\
\llap{$^a$}Tel Aviv University, Tel Aviv, Israel\\

E-mail: \email{levyaron@post.tau.ac.il}
 }

\abstract{
Following the decision to reduce the $L^*$ from 4.4~m to 4.1~m, the BeamCal had to be moved closer to the interaction point. Results of a study of how this affects the beamstrahlung and backward scattering backgrounds show that
the $e^+e^-$ pair background depositions from beamstrahlung at the BeamCal rises by 20\%.
The background from backscattered electrons and positrons in the inner pixel layers rises almost by a factor two, so does the number of photons in the tracker.}

\begin{document}

\section{Introduction}
\label{intro}

The physics program of the  planned future International Linear Collider (ILC)~\cite{ILC} is based on building two complementary detectors that will share beam time. For the sake of saving on the costs of the collider it was decided to use a push-pull concept to cost-effectively share the beam between the two detectors. The two detectors, the International Large Detector (ILD) and the Silicon Detector (SiD)~\cite{ILD-SiD} would be moving on slabs with pads or rollers. In their original design, the ILD had $L^*$ = 4.4~m and SiD, an $L^*$ = 3.5~m, where $L^*$ is the distance from the interaction point (IP) to the QD0 final focusing quadrupole magnet. By 2015 it was decided to adopt equal $L^*$ for both detectors, converging on one value of $L^*$ = 4.1~m. 

This talk presents results of a study to estimate the consequences of shortening $L^*$ for the ILD detector. After a description of the detectors in the ILD forward region, a possible solution for the new $L^*$ will be presented. A study of the backgrounds for this new setup  will include the background coming from beamstrahlung pairs and the one from backward scattering particles.

\section{The ILD forward region}

\begin{figure}[h!]
  \centering
   \includegraphics[width=0.7\textwidth] {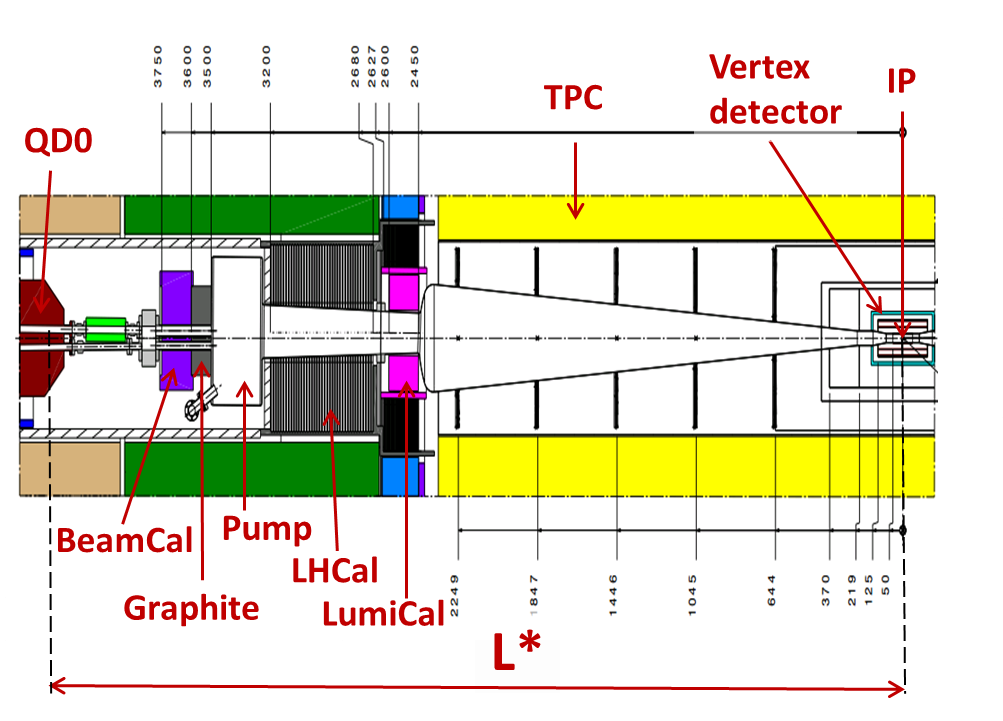}
\caption{The very forward region of the ILD detector. The Vertex detector surrounds the Interaction Point (IP) and is surrounded by the TPC.  In the forward direction LumiCal and BeamCal are installed, with LHCal and the vacuum pump in between.
A graphite ring is placed between the pump and BeamCal.
The distance between the IP and the QD0 magnet is denoted by $L^*$.}
\label{fig:forward-region}
\end{figure}

The forward region of ILD is shown in Fig.~\ref{fig:forward-region}. The Vertex detector surrounds the Interaction Point (IP) and is itself  surrounded by the TPC.  In the forward direction LumiCal and BeamCal~\cite{ilc1}  are installed, with LHCal and the vacuum pump in between. A graphite ring is placed between the pump and BeamCal as a shield of the latter.

The LumiCal is intended to measure the luminosity with a precision of better than 10$^{-3}$ at 500~GeV centre-of-mass 
energy and 3$\times 10^{-3}$ at 1~TeV centre-of-mass energy at the ILC, and with a precision of 10$^{-2}$ at CLIC~\cite{CLIC}. The BeamCal will perform 
a bunch-by-bunch estimate of the luminosity and, supplemented by 
a pair monitor, assist beam 
tuning when included in a fast feedback system~\cite{grah_sapronov}. The pair monitor is a layer of silicon pixel sensors designed to be located at the first layer of BeamCal and will be used for luminosity optimization.

\begin{figure}[h!]
  \centering
   \includegraphics[width=0.55\textwidth] {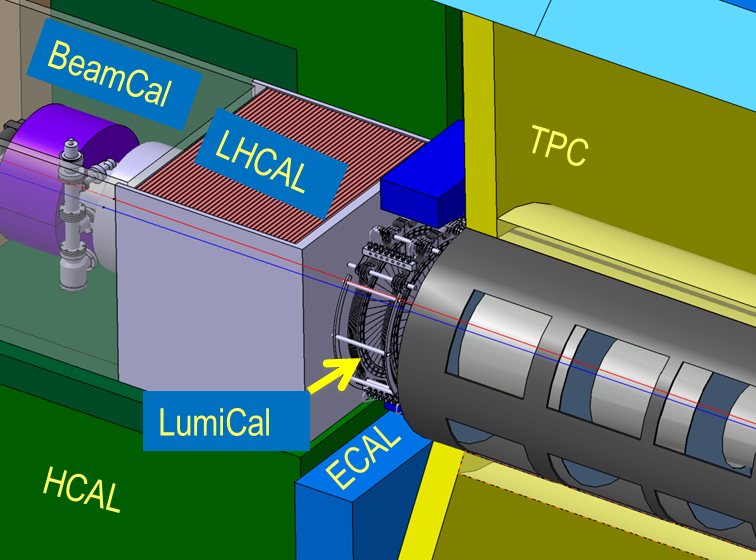}
\caption{The very forward region of the ILD detector. 
LumiCal, BeamCal and LHCAL are carried by 
the support tube for the final focusing quadrupole QD0 and the beam-pipe. 
TPC denotes the central tracking chamber, ECAL the electromagnetic and 
HCAL the hadron calorimeters. }
\label{fig:Forward_structure}
\end{figure}

LumiCal and BeamCal  extend the detector coverage to low polar angles, which is 
important e.g. for new particle searches with a missing energy signature~\cite{drugakov}. 
A sketch of the design is shown in Fig.~\ref{fig:Forward_structure} for the ILD detector. 
The LumiCal is positioned in a circular hole of the end-cap electromagnetic calorimeter ECAL.
The BeamCal is placed just in front of the final focus quadrupole magnet QD0.
LumiCal covers polar angles between 31 and 77~mrad and BeamCal, between 5 and 40~mrad.

The LHCal~\cite{Vlad} is a hadronic calorimeter in the forward direction which will help to separate electromagnetic and hadronic showers and will also extend the hermeticity of the ILD detector. 

\begin{figure}[h!]
  \centering
   \includegraphics[width=0.55\textwidth] {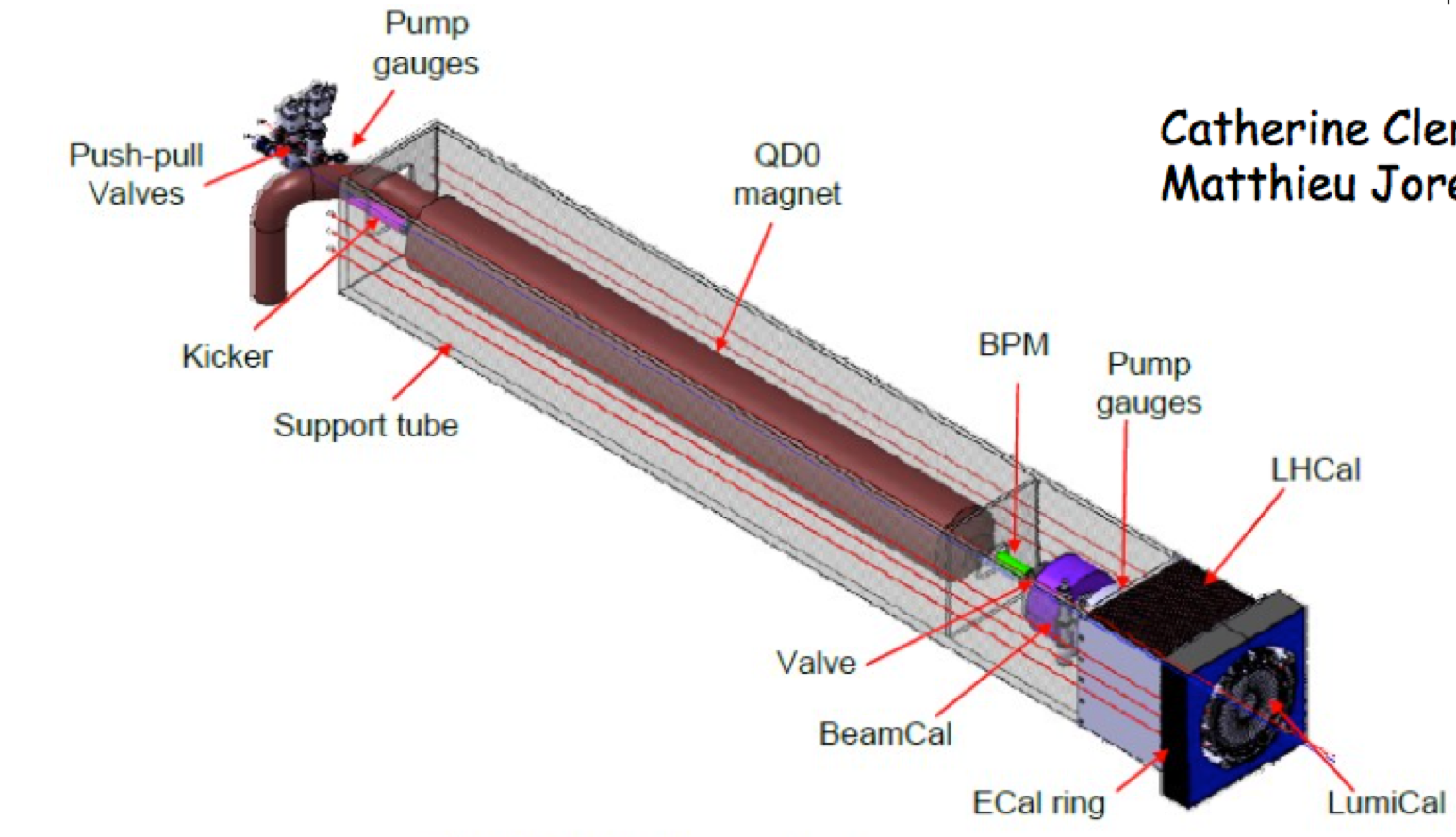}
\caption{The very forward region of the ILD detector, showing the beam-position monitor (BPM). All other elements are described in the captions of the earlier figures. 
 }
\label{fig:BPM}
\end{figure}

There is one more element in the very forward region, the beam-position monitor (BPM) to assist with the beam tuning (Fig.~\ref{fig:BPM}).

\section{Reduction of $L^*$ from 4.4~m to 4.1~m}

The ILC Baseline design, as described in the TDR~\cite{ILC}, is under revision since 2014. The proposed changes need to follow a defined procedure and need approval by the Linear Collider Collaboration (LCC) directorate. This process starts with a change request submitted as a document to the Change Management Board (CMB) and is being reviewed by experts. Their recommendations are then submitted to the ILC Director who makes, in consultation with the  CMB, the final decision.  This decision is transferred to the LCC directorate and if approved it is then released to be updated in the Technical Design Document.

In September 2014 a request was issued for changing the $L^*$ of both detectors to the same value of 4.1~m. It passed all the change-request procedure described above and was approved in October 2015. For ILD this meant moving the BeamCal closer to the IP. Since the BPM needs 10~cm, this meant that BeamCal had to be moved by 40~cm. 
\begin{figure}[h!]
  \centering
   \includegraphics[width=0.7\textwidth] {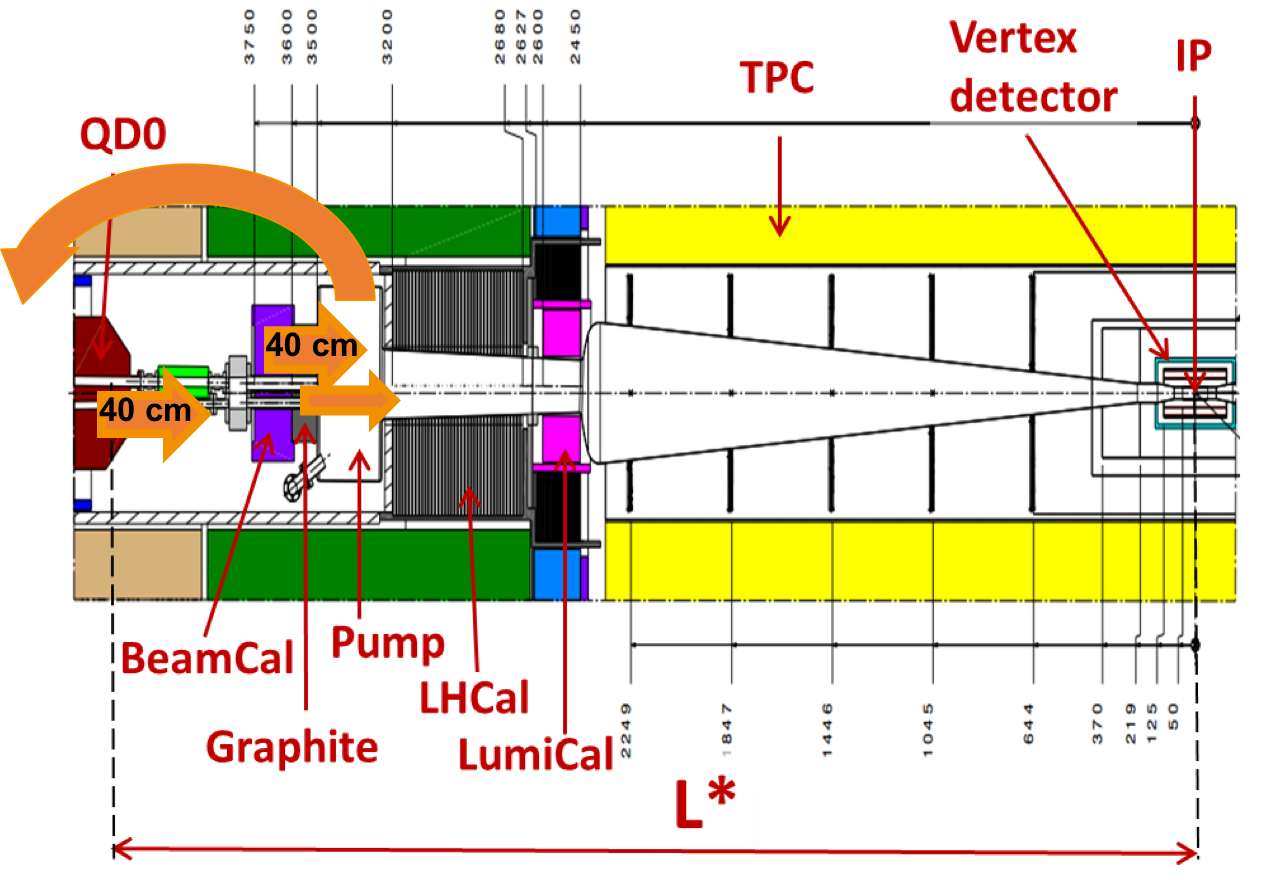}
\caption{The very forward region of the ILD detector, as in Fig.~\ref{fig:forward-region}, with arrows indicating the needed changes for a reduced $L^*$.
 }
\label{fig:Change-Lstar}
\end{figure}

The changes needed in ILD to cope with a smaller  $L^*$  are shown schematically in Fig.~\ref{fig:Change-Lstar}. A gain of 30~cm can be achieved by moving the vacuum-pump behind Q0. This can be done after preliminary studies showed that the expected changes in the  vacuum conditions and the impact of beam-gas scattering backgrounds are very small and acceptable~\cite{Karsten}. The additional 10~cm can be gained by redesigning LHCal such that the 10~cm graphite shield is moved into the LHCal inner cutout.

\begin{figure}[h!]
  \centering
   \includegraphics[width=0.7\textwidth] {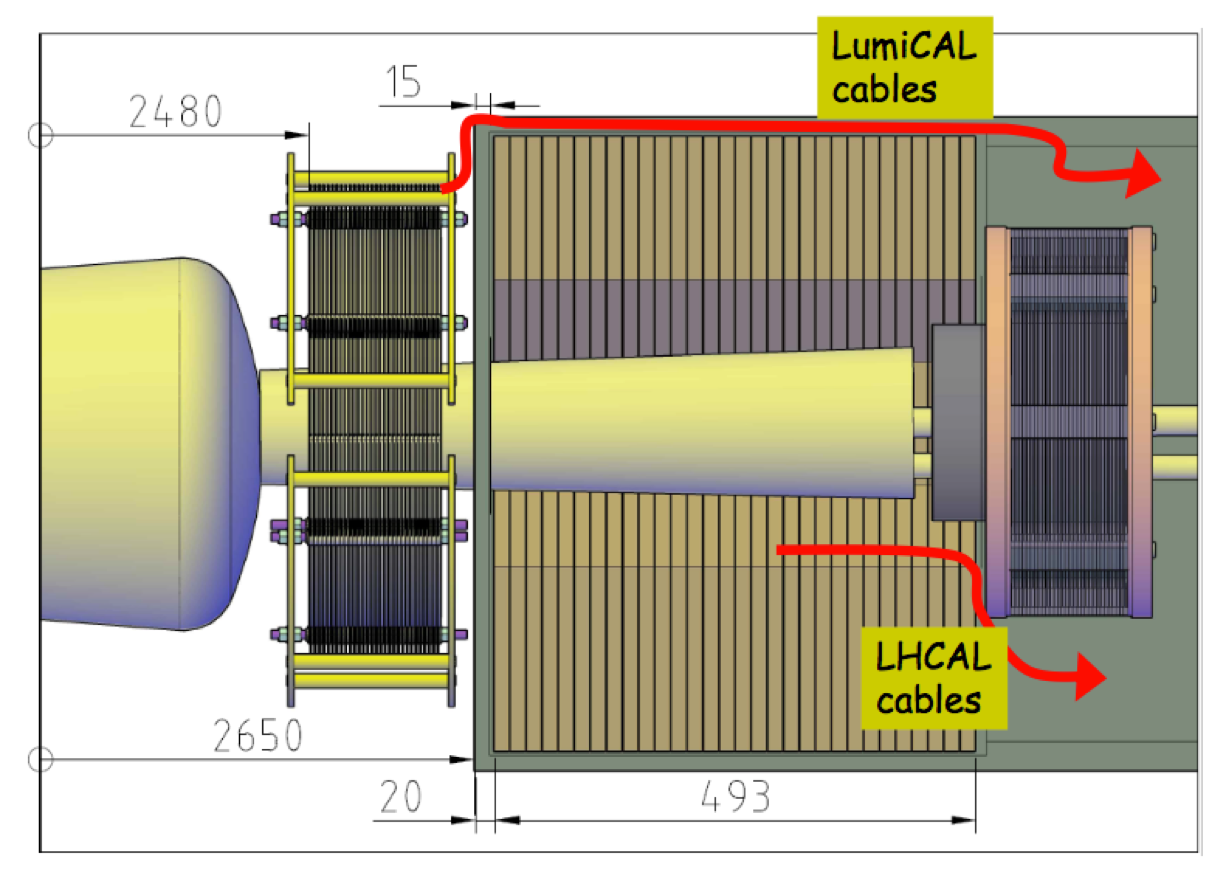}
\caption{A top view of the very forward region of the ILD detector with a reduced $L^*$, The BeamCal is moved by 40~cm closer to the IP. The red lines show the location of the cables from LumiCal and LHCal.
 }
\label{fig:Top-View}
\end{figure}

A top view of the new design of the very forward region of ILD can be seen in Fig.~\ref{fig:Top-View}, with the IP this time at the left-hand-side of the figure. The location of the cables from LumiCal and from LHCal are also indicated. The graphite shield is inside LHCal. Since BeamCal has been moved by 40~cm closer to the IP the second conical beam-pipe section gets shorter.

These changes have implications mainly on two issues. Having BeamCal closer to the IP means that keeping the original design of its inner radius (2~cm) would create empty space between the detector and the beam-pipe. In order to preserve the hermeticity function of BeamCal, its inner radius needs to be reduced to 1.78~cm.

\section{Background from beamstrahlung and backward scattering}

During the beam-beam interaction a large number of photons is produced, and then some of these photons produce $e^+e^−$ pairs. All of them together constitute beamstrahlung. In the magnetic field of the bunches,  the $e^+e^-$ pairs are deflected and hit the BeamCal. The flux of these pairs constitutes a background in the identification of single high energy electrons.

The other background affecting the ILD detector can come from backscattered particles, e.g. from the BeamCal, into other parts of the ILD detector. In the following, a study of these two backgrounds, after the BeamCal is moved to the new position, is presented. 

\subsection{Simulation}
\label{sec:geometry}

\begin{figure}[h!]
  \centering
   \includegraphics[width=0.5\textwidth] {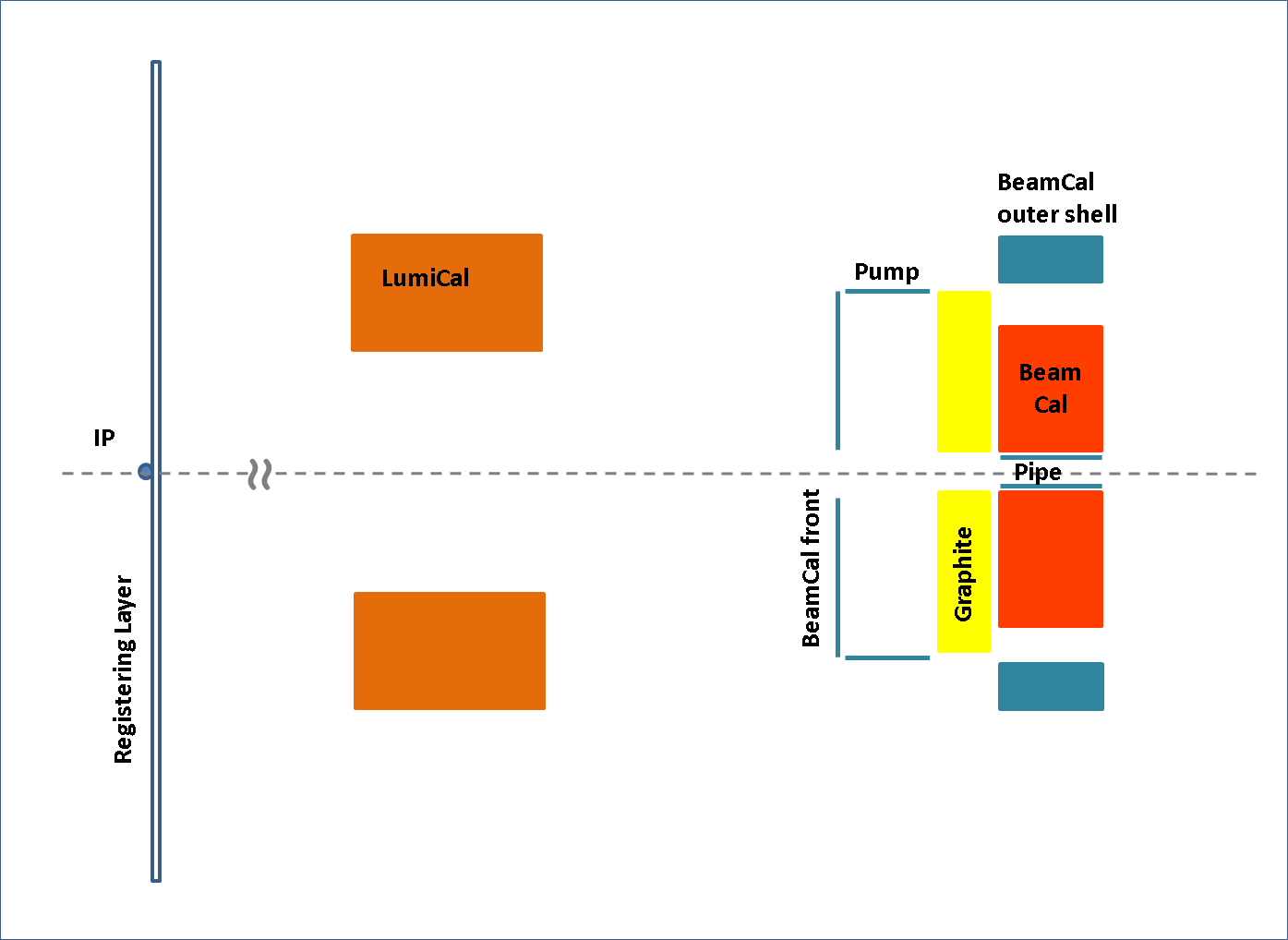}
\caption{A schematic sketch of the simulation geometry used for studying the background from BeamCal.}
\label{fig:simulation-model}
\end{figure}

In order to study the influence of the $L^*$ change on the backgrounds coming from beamstrahlung and backscattering, the {\sc GEANT4}~\cite{geant4} based simulation package {\sc BeCaS}~\cite{becas} has been used (Fig.~\ref{fig:simulation-model}).  
The package includes an anti-DID map of the magnetic field, pipes, support, the graphite filter and a rough LumiCal model.   It performs shower development in BeamCal and creates backscattered particles. The backscattered particles are probed at a plane perpendicular to the beam,  at the position of the IP.

\begin{figure}[h!]
  \centering
   \includegraphics[width=0.4\textwidth] {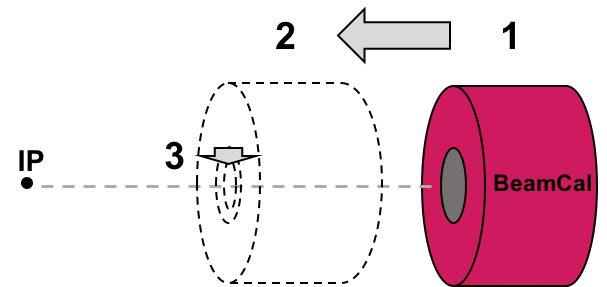}
\caption{A schematic sketch of three different geometries of BeamCal.}
\label{fig:Geometries}
\end{figure}

Three different geometries, as schematically shown in Fig.~\ref{fig:Geometries}, were considered.
\begin{itemize}
\item
{\bf Geometry 1}:
The original baseline design. The distance of BeamCal from the IP is 360~cm, and the inner radius of BeamCal, $R_{in}$, is 2~cm.
\item
{\bf Geometry 2}:
BeamCal distance to the IP is 320~cm, $R_{in}$ is 2~cm.
\item
{\bf Geometry 3}:
BeamCal distance to the IP is 320~cm, $R_{in}$ is 1.78~cm in order to cover the same lower polar angle as for the baseline design.
\end{itemize}

\subsection{Beamstrahlung pairs}

\begin{figure}[h!]
  \centering
   \includegraphics[width=0.75\textwidth] {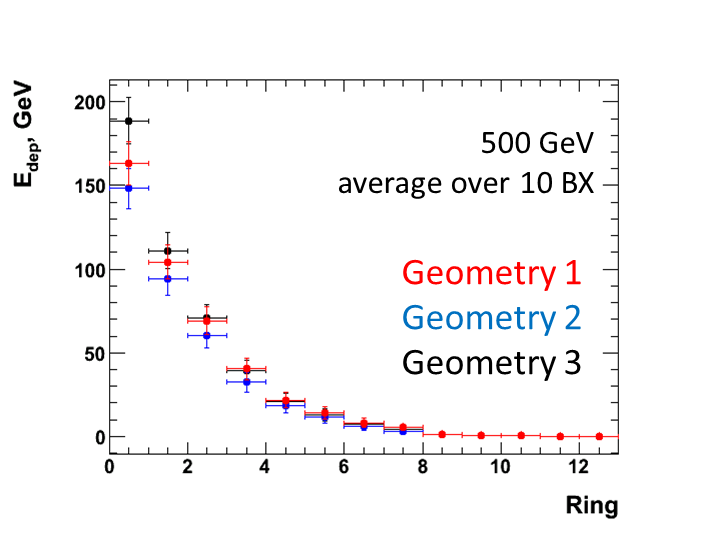}
\caption{Energy deposited in BeamCal by beamstrahlung pairs, $E_{dep}$, as a function of the ring number of the BeamCal, for 500~GeV incoming beams. The energy is averaged over 10 bunch-crossings (BX) and displayed for Geometries 1 (red), 2 (blue) and 3 (black).}
\label{fig:E-beamstr}
\end{figure}

The beamstrahlung pairs were traced through the solenoidal 4T magnetic field of ILD with anti-DID correction and the energy deposited in each ring of BeamCal was investigated. The study was performed for the three geometries defined in Section~\ref{sec:geometry}, for the 500~GeV  option of the ILC. Figure~\ref{fig:E-beamstr} shows the energy deposited in BeamCal by beamstrahlung pairs, $E_{dep}$, as a function of the ring number of the BeamCal. The energy is averaged over 10 bunch-crossings (BX) and displayed for the three geometries.
Much of the energy of the beamstrahlung pairs is deposited in the first two rings. When the BeamCal moves closer to the IP and keeps its original inner radius of $R_{in}$ = 2~cm, it covers bigger outer polar angles but some smaller polar angles remain uncovered and thus particles from small angle beamstrahlung pairs remain in the beam-pipe and do not deposit energy in BeamCal. This can be seen in the decrease of deposited energy for Geometry 2.  When, additionally, also $R_{in}$ is reduced to 1.78~cm, BeamCal covers a bigger range of polar angles, from the original polar angle of Geometry 1 to the outer angle of Geometry 2. As a results the energy deposited per ring is bigger compared to that of the baseline geometry. A similar behaviour is also seen for the 1~TeV option (not shown) with the energy deposition per ring approximately three times larger than for the 500~GeV option.  

The conclusion from this study is that the pair background from beamstrahung increases by about 20\% by moving the BeamCal closer to the IP and by decreasing its inner radius for the sake of hermeticity.

\subsection{Backscattered particles}

\subsubsection{Geometry 1}

\begin{figure}[h!]
  \centering
   \includegraphics[width=0.45\textwidth] {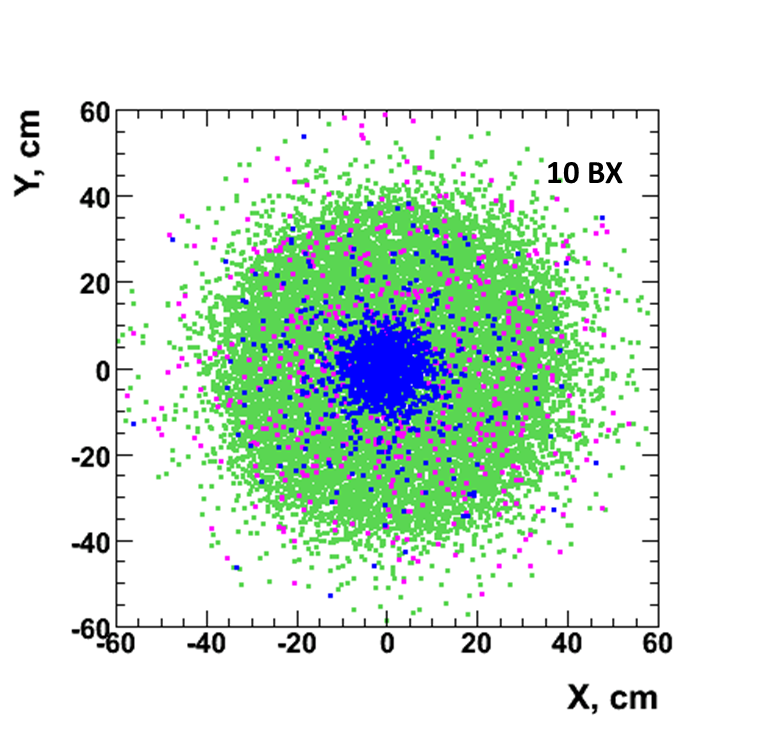}
\caption{Distribution of hits in the $X$-$Y$ plane of the Registering Layers coming from all backscattered particles. Electrons and positrons (blue), photons (green) and neutrons (pink).}
\label{fig:all-back}
\end{figure}

\begin{figure}[h!]
  \centering
   \includegraphics[width=0.32\textwidth] {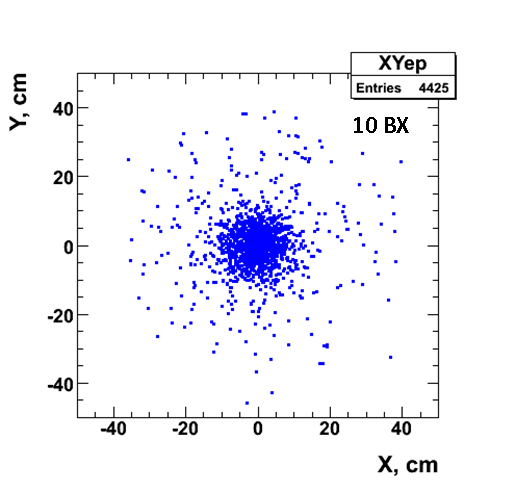}
\includegraphics[width=0.32\textwidth] {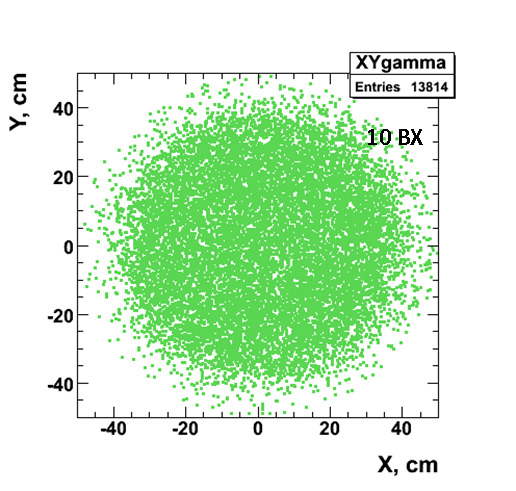}
\includegraphics[width=0.32\textwidth] {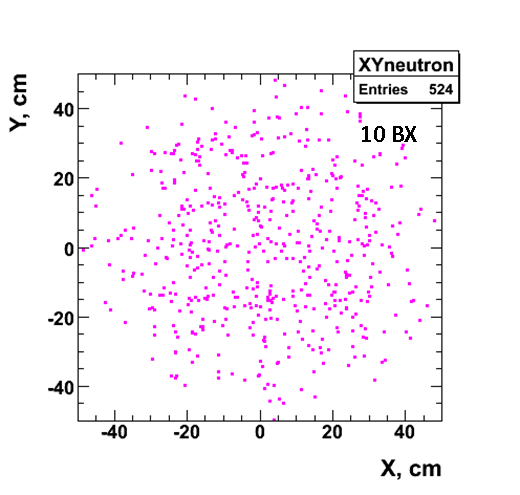}
\caption{Distribution of hits in the $X$-$Y$ plane of the Registering Layer coming from backscattered particles. Left: electrons and positrons. Center: photons. Right: neutrons.}
\label{fig:single-back}
\end{figure}

A review of the characteristics of the backscattered particles for the original geometry as in the baseline design, namely Geometry 1, is first presented. To this end, the distribution of particles backscattered onto the Registering Layer (see Fig.~\ref{fig:simulation-model}) in the transverse plane to the beam direction is described. In a reference system where the beam is along the $Z$ axis, the hits of backscattered particles in the $X$-$Y$ plane are shown in Fig.~\ref{fig:all-back} for all particles that are backscattered in 10 BX.
Almost all hits are within a radius of 50~cm, mainly due to the screening by the LumiCal. The different colors indicate the types of particles. The blue color concentrated at the origin (beam-pipe) are electrons or positrons. The green color distributed uniformly are photons and the pink color, also uniformly distributed, are neutrons. This is shown separately in Fig.~\ref{fig:single-back}.

The composition of backscattered particles per BX can be quantified by averaging the number of particles in 10 BX. The numbers are presented in Table~\ref{tab:composition}. Close to three quarters of the backscattered particles are photons. The  electrons and positrons constitute about 23\%, the neutrons about 3\%, with a negligible number of other charged particles.
\begin{table}[h]
\centering
  \begin{tabular}{ | c | c | c |}
    \hline    
	Particle type		&  Number of particles      &  Percentage  \\ \hline \hline
$\gamma$		&	1381.4	& 	73.4 \%	\\ \hline
$e^-$/$e^+$			&   	373.7/68.8	& 	19.8 \%/3.7\%	 \\ \hline
n                        &  52.4 & 2.8\% \\ \hline
$\mu^\pm$, $\pi^\pm$, p   &   4.9 & 0.3\% \\ \hline
  \end{tabular}
  \caption{The composition of backscattered particles per bunch crossing, averaged over 10 BX, in a circular area of the Registering Layer within a 50~cm radius for Geometry 1, for a 500~GeV ILC.}
\label{tab:composition}
\centering
\end{table}

A similar study for a 1~TeV ILC gives a total number of backscattered particles about three times larger, with a similar distribution of the types of particles.

\begin{figure}[h!]
  \centering
   \includegraphics[width=0.45\textwidth] {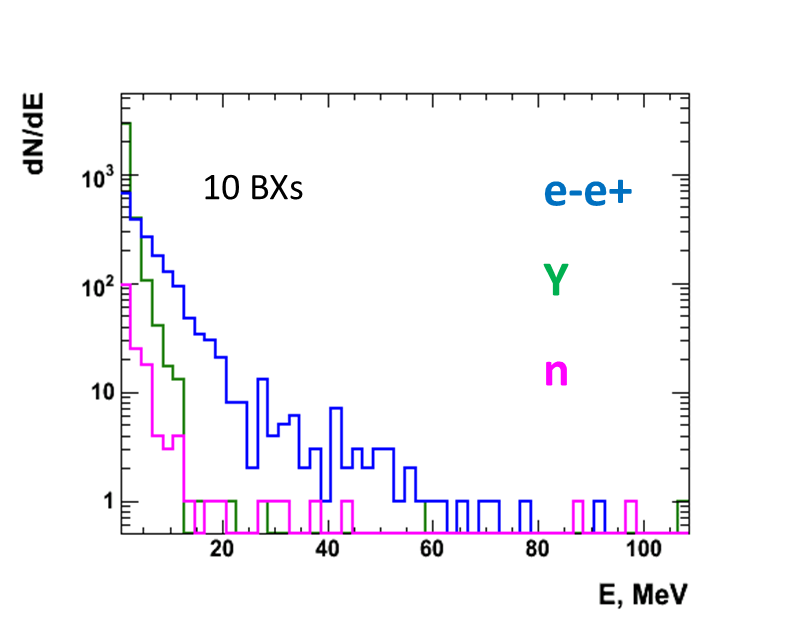}
\includegraphics[width=0.45\textwidth] {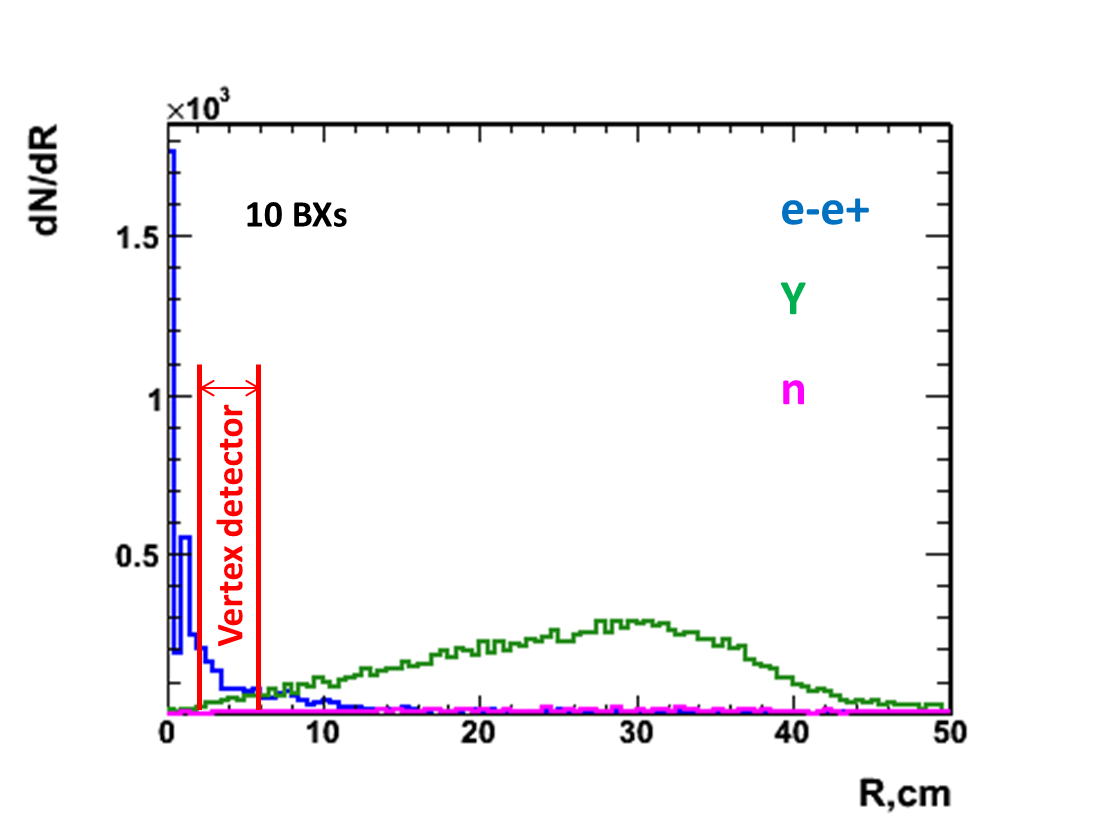}
\caption{Distribution of backscattered particles in 10 BX for electrons or positrons (blue), photons (green) and neutrons (pink), for Geometry 1. Left: Energy, Right: Radial. The two red lines show the position of the vertex detector.}
\label{fig:ER-back}
\end{figure}

The energy and radial distribution of the backscattered particles are presented in Fig.~\ref{fig:ER-back}.
All particle types are distributed over a similar range of energy, extending up to several tens of MeV. A significant part of the $e^-$ and $e^+$ particles remain in the beam pipe. However some of them hit the vertex detector, especially the inner layer.

\subsubsection{Different Geometries}

\begin{figure}[h!]
  \centering
   \includegraphics[width=0.45\textwidth] {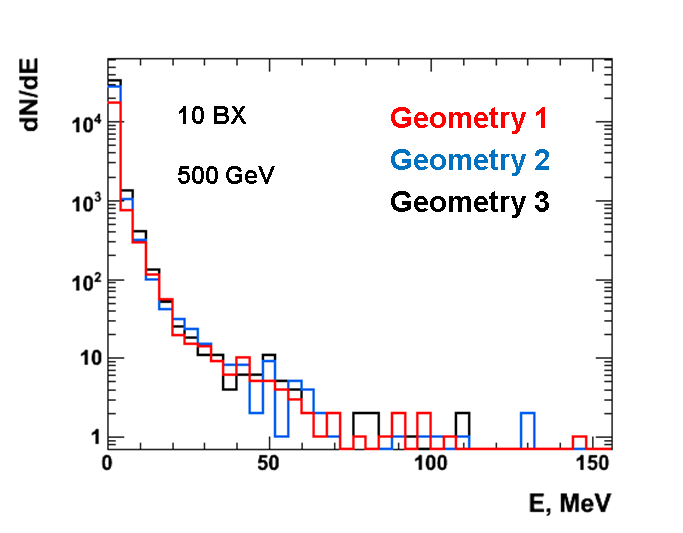}
\includegraphics[width=0.45\textwidth] {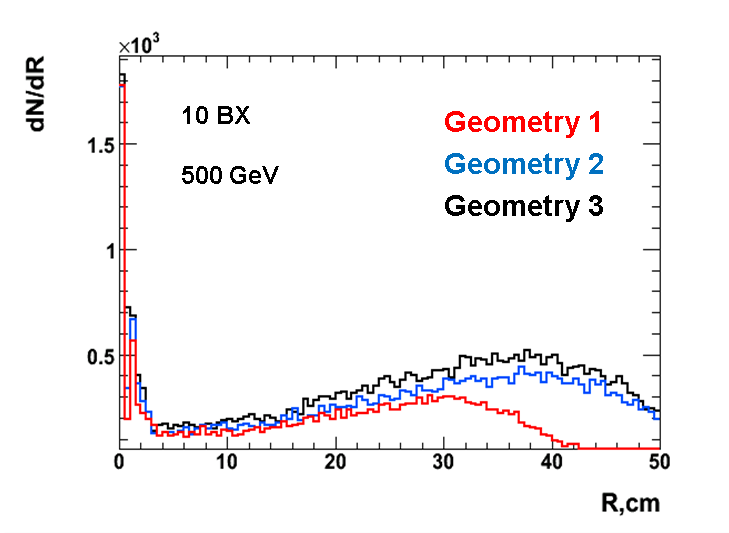}
\caption{ Distribution of backscattered particles in 10 BX for Geometry 1 (red), Geometry 2 (blue) and Geometry 3 (black). Left: Energy, Right: Radial. }
\label{fig:geo-back}
\end{figure}

The energy and radial distribution of backscattered particles have been studied separately for the three geometries mentioned above, and are shown in Fig.~\ref{fig:geo-back}.
The number of particles increases from Geometry 1 to 3, but the energy distributions do not differ significantly. The number of particles at larger radii  increases, mostly due to a larger number of photons at large radii.

\subsubsection{Occupancy in the ILD vertex detector}

\begin{figure}[h!]
  \centering
   \includegraphics[width=0.95\textwidth] {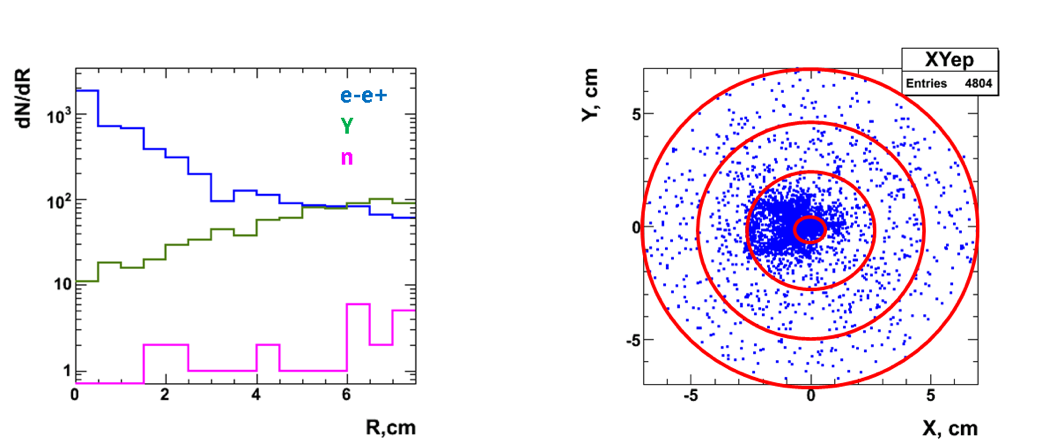}
\caption{Left: The distribution of backscattered particles from 10 BX into the vertex detector, as a function of the radius. Right: the distribution of the electron and positrons in the $X$-$Y$ plane.}
\label{fig:occupancy-vtx}
\end{figure}

The ILD vertex detector is hit by beamstrahlung and by backscattered particles even in the baseline design. Here,  the increase of the backscattered particles due to the move of BeamCal towards the IP, for the case of Geometry 3, is presented. The vertex detector is a multi-layer pixel detector, consisting of three cylindrical concentric double layers at radii from 16 to 60~mm. To study the occupancy of backscattered particles  at the IP, it is divided into rings, such that the boundary radii are in the middle between vertex detector double layers. The radial distribution of the backscattered particles from 10 BX into the vertex detector is shown in Fig.~\ref{fig:occupancy-vtx} and as the charged particles mostly affect the vertex detector, the distribution of the electrons and positrons in the $X$-$Y$ plane is also shown in the figure. The latter has overlayed circles that show the three rings of the vertex detector.  
The photons and neutrons are distributed uniformly (not shown). The electrons and positrons are mostly concentrated near the beam axis and have an asymmetrical shape. The largest occupancy is in the first layer of the vertex detector.

\begin{table}[h!]
\centering
  \begin{tabular}{ | c | c | c | c | c | c |}
    \hline    
	{\Large{$e^-, \ e^+$}	}	&  Pipe     &  Layer 1 & Layer 2 & Layer 3  & SIT, TPC\\ \hline \hline
Geometry 1		&	180.8	& 	123.9 & 40.4 & 26.6 & 70.8	\\ \hline
Geometry 2		&   	181.1	& 	164.8 (+33\%) & 41.3 (+2\%) & 27.7 (+4\%) & 79.3 (+12\%)	 \\ \hline
Geometry 3              &     195.1      &      205.3 (+66\%) & 46.4 (+15\%) & 33.2 (+25\%) & 85.6 (+21\%)  \\ \hline
	{\Large{$\gamma$}	}	&    &  & &   & \\ \hline \hline
Geometry 1		&	1.2	& 	6.1& 40.4 & 17.2 & 1333.8	\\ \hline
Geometry 2		&  1.1	& 	8.1 (+33\%) & 15.3 (-11\%) & 26.0 (+12\%) & 2336.9 (+75\%)	 \\ \hline
Geometry 3              &     1.7      &     9.3 (+52\%) & 29.3 (+18\%) & 35.2 (+52\%) & 2798.8 (+110\%)  \\ \hline
	{\Large{$n$}	}	&    &  & &   & \\ \hline \hline
Geometry 1		&	0.0	& 	0.3& 0.7 & 0.8 & 50.6	\\ \hline
Geometry 2		&  0.1	& 	0.5 (+60\%) & 0.6 (-15\%) & 1.3 (+62\%) & 68.9 (+36\%)	 \\ \hline
Geometry 3              &  0.0     &     0.4 (+30\%) & 0.6 (-15\%) & 0.9 (+12\%) & 88.6 (+75\%)  \\ \hline
  \end{tabular}
  \caption{The average number of backscattered particles for 1 BX for each ring  of the vertex detector, per Geometry, for electrons or positrons, for photons and for neutrons. In brackets, the percentage change compared to Geometry 1 is given. The last column gives the number of particles in the SIT and TPC detectors. }
\label{tab:occupancy}
\centering
\end{table}
In order to quantify the increase of the occupancy due to the change in Geometry,  Table~\ref{tab:occupancy} lists the average number of backscattered particles for 1 BX for each ring, per Geometry. For Geometries 2 and 3, the  percentage change of the number of particles  relative to Geometry 1 is also given. In the last column of the Table, the number of backscattered particles into the SIT and TPC detectors are shown. The SIT is the intermediate silicon-strip barrel detector with two double layers bridging the gap between the vertex detector and the TPC.

The vertex detector is  sensitive to   electrons and positrons. Their amount increases  from Geometry 1 to Geometry 3 and the first layer shows the largest increase of 66 \%.

\section{Conclusions} 
The design of the ILD forward region was revisited to match the new $L^*$ = 4.1~m.
The BeamCal was shifted by 40~cm towards the IP direction, with the
vacuum pump moved behind the focusing quadrupole QD0 magnet. The 
graphite filter absorber was placed inside the LHCal inner cutout.

The influence of the change of $L^*$ was simulated and the deposited energy in the BeamCal and backscattered particles in the central region of the ILD was studied.
The $e^+e^-$ pair background depositions from beamstrahlung at the BeamCal rises by 20\%.
The background from backscattered electrons and positrons in the inner pixel layers rises almost by a factor 2, so does the number of photons in the tracker. Since all these results are based on a stand-alone MC simulation, it should be repeated with a study using the DD4hep MC model~\cite{dd4hep}  of forward region.

\section*{Acknowledgments}
This study was partly supported by the Israel Science Foundation (ISF), Israel German Foundation (GIF), the I-CORE program of the Israel Planning and Budgeting   Committee, Israel Academy of Sciences and Humanities, 
and by the European Union Horizon 2020 Research and Innovation programme under Grant Agreement no.654168 (AIDA-2020).

\end{document}